\documentclass[11pt]{article}
\usepackage{cite,graphicx,graphics,epsfig}
\newcommand{\ba}{\begin{array}}
\newcommand{\ea}{\end{array}}
\newcommand{\bd}{\begin{displaymath}}
\newcommand{\ed}{\end{displaymath}}
\newcommand{\be}{\begin{equation}}
\newcommand{\ee}{\end{equation}}
\def\bt{\begin{table}}
\def\et{\end{table}}
\def\bi{\begin{itemize}}
\def\ei{\end{itemize}}
\def\bea{\begin{eqnarray}}
\def\eea{\end{eqnarray}}

\def\a{\alpha}

\def\g{\gamma}
\def\d{\delta}

\def\G{\Gamma}

\def\s{\sigma}

\def\q2 {q^2}

\def\r {\rightarrow}

\catcode`@=11 
\def \gsim{\mathrel{\mathpalette\@versim>}}
\def \lsim{\mathrel{\mathpalette\@versim<}}
\def \@versim#1#2{\lower0.4ex\vbox{\baselineskip\z@skip\lineskip\z@skip
     \lineskiplimit\z@\ialign{$\m@th#1\hfil##\hfil$%
     \crcr#2\crcr\sim\crcr}}}
\catcode`@=12 
\begin{document}
\begin{center}
{\large\sc  Effects of Universal Extra Dimensions on Higgs signals at 
LHC }\\[5mm]
Santosh Kumar Rai\footnote{E-mail: santosh.rai@helsinki.fi}\\
{\em  High Energy Physics Division, Department of Physical Sciences, University
     of Helsinki,\\ and Helsinki Institute of Physics, P.O. Box 64,
     FIN-00014 University of Helsinki, Finland}
\end{center}
 \vskip 10 mm

\begin{abstract}
\noindent
A major focus at the Large Hadron Collider (LHC) will be on Higgs boson 
studies and it would be an interesting prospect to simultaneously probe for 
physics beyond the Standard Model (SM) in the Higgs signals. In this work
we show as to what extent, the effects of Universal Extra Dimension (UED)
can be isolated at the LHC through the Higgs signals. By doing a detailed 
study of the different uncertainties involved in the measurement of the 
rates for the process $pp \to h \to \g\g$ we estimate the extent to which 
these uncertainties can mask the effects of the contributions coming 
from UED.
\end{abstract}
\vskip 0.2 true cm


\section{Introduction}
The much anticipated experiments at the Large Hadron Collider (LHC) 
are expected to refine our understandings of the Standard Model (SM)
further and also shed some light on physics beyond the SM. But most 
importantly, the LHC is envisioned as the machine to complete the 
picture of the SM by discovering the Higgs boson, or instead, give a 
hint into the mechanism responsible for electroweak symmetry breaking. 
Being a subject of so much speculation and study in particular, one would 
always want to find out if the study on the Higgs sector itself may 
reflect information on any kind of new physics beyond the SM. 
In this work, we explore this above possibility by studying the effects 
of new physics on the signal of Higgs boson in the light of experiments 
at the LHC.

Within some models of extradimensions motivated from the framework 
of string theory \cite{antoniadis1}, non-gravitational fields are also 
free to propagate in the bulk provided they do not disturb the 
experimental constraints. One such scenario, referred to as the 
Universal Extra Dimension (UED) model \cite{appel}, allows all the SM 
fields to propagate in the extra dimension. In the effective four 
dimensional space-time the effects of the extra dimension is felt through 
the Kaluza-Klein (KK) excitations of these bulk fields which interact with
the SM particles (identified as the zero modes of the excitations).
At tree level, the momentum along the extra dimensions is conserved, 
which requires pair production of these Kaluza-Klein (KK) modes at 
colliders and preventing tree level mixing effects from altering 
precision electroweak measurements. 
Values of the compactification scale are constrained, and it
has a lower bound of about 300 GeV \cite{appel-yee}. 
The phenomenological implications of UED have been extensively studied in 
the literature \cite{appel,phenolow1,phenolow2,phenolow3,phenolow4,phenolow5,phenolow6,phenolow7,phenohigh1,phenohigh2,phenohigh3,phenohigh4,phenohigh5,phenohigh6,phenohigh7,phenohigh8,phenohigh9,phenohigh10,phenohigh11,phenohigh12,phenohigh13,phenohigh14}.

Direct detection of UED KK states at future colliders requires them to 
be pair produced due to the KK number conservation and hence already puts 
a limit on the minimum energy at which the collider should run to produce 
these particles. The LHC will invariably be able to probe physics at the 
energy regime unconstrained by precision measurements, where such particle 
resonances are expected to occur. 
The main experimental signal for the 
production and decay of KK excitations at hadron colliders will be the 
observation of events with multiple leptons and jets of moderately high 
energies in association with large missing energy \cite{Macesanu:2005jx}. 
This draws a lot of parallels with supersymmetric searches at the hadron 
colliders and it will prove to be a strong challenge to distinguish the 
signatures. We refer the readers to \cite{Macesanu:2005jx} where one gets a 
nice review of different interesting signatures at colliders. 
However, it would be worthwhile to look 
for its effects in Higgs boson studies whose signals would be extensively 
studied. There would be possible modifications in the signal, through the 
modification of Higgs decay properties due to the KK states contributing 
in the loop mediated decay modes of the Higgs boson. The partial decay widths 
for $h\to gg$, $h\to\g\g$ and $h\to\g Z$ decay modes which are driven by 
loops can be substantially modified due to KK excited modes of SM particles 
running in the loops.  There is in fact  
remarkably significant enhancement in the partial decay width of the Higgs 
in $h\to gg$ due to the excited top quark loops \cite{petriello}. This can
greatly enhance the Higgs production at the LHC viz. the $gg \to h$ mode 
of production. The production mode $ gg \to h \to \g\g$ is relevant for 
the Higgs boson lying in the mass range of $120-150$ GeV. Due to the 
limitations in the resolution of the calorimeter the measurement of the 
decay width of the Higgs boson in this mass range will be impossible.
Thus it would be impossible to study the partial widths $\G_{gg},
\G_{Z\g}$ and $\G_{\g\g}$ and look for any kind of UED effects. It 
would require study of event rates for its production and try to extract the 
contributions of new physics through the analysis of the rates for the
above mentioned process. A study considering rates to identify UED 
effects in Higgs signals at a linear $e^+ e^-$ collider has been 
looked into, in ref~\cite{skrai}.  

However, the effects would be masked by the different uncertainties
which will affect measurements at the LHC and thus make it difficult 
to differentiate the contributions coming from UED. In this work we 
look at the dominant mode of Higgs production 
through the $gg$-fusion for a Higgs in the mass range of $120-150$ GeV
and its subsequent decay into two photons and try to identify the 
contribution coming from UED and the extent to which these can be
identified over the uncertainties that would affect measurements at 
the LHC. A very similar analysis has been recently carried out, in
context of Split Supersymmetry, in identifying additional contributions to
the Higgs rate \cite{Gupta}. In Section~2 we give a very brief overview 
about the model in consideration. In Section~3 we discuss the process
under consideration and how the signals for the diphoton final states 
get modified due to UED contributions. In Section~4 we discuss the
different uncertainties that would affect the signals. In
Section~5 we present our numerical results and finally we summarise and 
conclude in Section~6.

\section{The Minimal Model}
The UED model, in its simplest form \cite{appel}, has all the SM
particles propagating in a single extra dimension, which is
compactified on an $S_1/Z_2$ orbifold with $R$ as the radius of
compactification.  Conservation of KK number which is a consequence of 
momentum conservation along the extra dimension forces the KK particles 
to be pair produced. Consequently, UED predicts a stable lightest 
Kaluza-Klein particle (LKP) which would be much like the lightest 
supersymmetric particle (LSP) and a prospective candidate for dark
matter \cite{LKP1,LKP2}. Bulk and brane radiative 
effects \cite{branebulk1,branebulk2,branebulk3,branebulk4} 
however break KK number down to a discrete conserved quantity, the so 
called KK parity, $(-1)^n$, where $n$ is the KK level. KK parity 
conservation in turn, implies that the contributions to various 
precisely measured low-energy observables only arise at loop level and 
are small \cite{phenolow1,phenolow2,phenolow3,phenolow4,phenolow5,phenolow6,phenolow7}. 

The KK tower resulting on the four dimensional space-time has a tree
level mass given by
\bea
 m_n^2 = m^2 + \frac{n^2}{R^2}
\eea
where $n$ denotes the $n^{th}$-level of the KK tower and $m$
corresponds to the mass of the SM particle in question. This implies a
mass degeneracy in the $n^{th}$-level of the spectrum at least for the
leptons and lighter quarks. This degeneracy is however
removed due to radiative corrections to the masses \cite{branebulk1,branebulk2,branebulk3,branebulk4}.

\section{Higgs signals and the diphoton mode}
If the Higgs exists in the mass range of $120-150$ GeV, then we 
should be able to  see it during an early phase of the LHC. 
If that is possible, then it will be interesting to see if there are
any indications of new physics in Higgs signal itself, even if 
the detection of any new particle beyond the SM might not be possible
due to their high mass. 

The most suggestive channel in this context, for a Higgs boson in the
mass range $120-150$ GeV, is the  production of the Higgs through the
$gg$-fusion channel followed by its decay into the diphotons. In this 
mode, the (partial) decay width $\Gamma(h \r g g)$, gets additional 
contributions from the KK excitation of the top quark, while the
(partial) decay width $\Gamma(h \r \g \g)$ gets additional
contributions from the KK excitation of both the top quark and the
$W-$boson alongwith its associated Goldstone modes, ghost KK states, 
and in addition, also due to the charged Higgs tower. It has been 
shown in quite detail \cite{petriello} that these loop contributions 
alter the Higgs decay widths, thus making it distinguishable from the 
SM Higgs boson. 
In this work, we are mainly interested in the modification of these
partial decay width of the Higgs. Since the KK number is not violated at 
any of the vertices inside a loop, the contributions come from all the 
KK-excitations, with a decoupling nature for the higher modes. 
The combined expressions for the partial decay width for $h \to gg$ and 
$h \to \g\g$ for both UED and SM contributions can be written down as,
\bea
\G(h\to gg) = \frac{G_F~m_h^3}{36\sqrt{2}\pi}
\left(\frac{\alpha_s(m_h)}{\pi}\right)^2~|I_q+\sum_n {\tilde
I}_{q^{(n)}}|^2 \\
\G(h \to \g\g) = \frac{G_F}{128\sqrt{2}}
\frac{\alpha^2_{em} m_h^3}{\pi^3}~|I_q + I_W + 
\sum_n {\tilde I}_{q^{(n)}} + \sum_n {\tilde I}_{W^{(n)}}|^2
\eea
where $G_F$ is the Fermi constant, $\alpha_s(m_h)$ is the running
QCD coupling evaluated at $m_h$, $\alpha_{em}$ is the electromagnetic
coupling  and $I_i, {\tilde    I}_{i^{(n)}}$ are the
contributions of the loop integrals for the SM and UED case
respectively. We consider the contributions from the KK 
excitation of the top quark as well as the bottom quark as we wish to
make precise estimates comparable to uncertainties.  
We have to include the KK excitations of the $W-$boson and its
associated Goldstone modes, ghost KK states and the charged Higgs tower
for the diphoton decay channel. 
The UED contributions include the sum over the KK towers of the
respective particle. As the more massive modes in the loop will hardly
make significant contributions, we ensure that the sum is terminated 
as the higher modes decouple. We include all the decay modes affected by 
UED contributions in the decay package HDECAY \cite{hdecay} to evaluate the 
relative sensitivities to the branching ratios to the different decay channels
of the Higgs boson. 

It has to be remembered, however, that the above decay width will not be
a directly measurable quantity at the LHC. This is because the width is 
of the order of keV in the relevant Higgs mass range, which is smaller 
than the resolution of the electromagnetic calorimeters to be used
\cite{atlas, cms}. Here we try to estimate how
the UED contributions may be extracted in this channel, given the rather
sizable theoretical as well as experimental uncertainties in the 
various relevant parameters.

We, therefore, have chosen to do a  calculation involving the 
full process $(p p \r h X \r \g \g)$, that is to say, the
production of the Higgs followed by its decay into the diphoton final state.
Taking all uncertainties into account, we have tried to find the
significance level at which the additional contributions can be
differentiated in different regions of the parameter space which in
this scenario is the compactification radius $R$.
We have confined ourselves to the production of Higgs via gluon fusion.
The other important channel, namely gauge boson fusion, has been left 
out of this study, partly because it is plagued with uncertainties arising,
for example, from diffractive production, which may be too large for
the small effects under consideration here. 
In the SM, the loop-induced decay widths of the Higgs boson, including 
QCD as well as further electroweak corrections, are well-documented in 
the literature \cite{qcdcorr1,qcdcorr2,qcdcorr3,qcdcorr4,qcdcorr5,qcdcorr6,qcdcorr7,qcdcorr8,qcdcorr9,qcdcorr10,2lewc1,2lewc2}.

The rate for the inclusive process  
$$ p p \r h ~+~ X \longrightarrow \g \g$$ 
(where Higgs production takes place via gluon fusion) can be expressed 
in the leading order as
\begin{equation}
N~=~ \frac{\pi^2}{8 m_h s} \frac{\G_{h \r 2g} \G_{h \r 2\g}}
{\G_{tot}}\int^{1}_{\tau} {d\zeta \frac{1}{\zeta} 
g\left(\zeta,m^2_h\right)~g\left(\frac{\tau}{\zeta},m^2_h\right)}
\end{equation}
where~ $\tau =\frac{m^2_h}{S}$ and $g\left(\zeta,m^2_h\right)$
is the gluon distribution function evaluated at $Q^2~=~m^2_h$ and parton
momentum fraction $\zeta$. $ \G_{h \r 2\g}$, $ \G_{h \r 2g}$ and 
$\G_{tot}$ stand respectively for the diphoton, two-gluon and total 
decay widths of the Higgs.
The lowest order estimate given above is further multiplied by the 
appropriate K-factors to obtain the next-to-next leading order (NNLO)
predictions in QCD. While the computation of the rate is straightforward, 
we realise that the various quantities used are beset with 
theoretical as well as experimental uncertainties
\cite{zepnew,pdfscale,expnos}. We undertake an analysis of these 
uncertainties in the next section.

\section{Numerical estimate: uncertainties} 
As has already been stated in the previous section, 
the rate for diphoton production through real Higgs at LHC is given by
\bea
N~=~ \s(p p \r h) \times B~=~\s(p p \r h)
{\frac {\Gamma(h\r\gamma\gamma)}{\Gamma_{tot}}}
\eea
We have performed a parton-level Monte Carlo calculation for the production 
cross-section, using  the MRS~\cite{mrst} parton distribution functions
and multiplied the results with the corresponding  NNLO K-factors
\cite{kfactor, hocorrs1,hocorrs2}.
It may be noted that NNLO K-factors are not yet available for most
other parameterizations. In estimating the statistical uncertainties in 
the experimental value \cite{expnos}, MRS (at leading order) 
distributions have been used by the CMS group while  ATLAS uses CTEQ 
distributions. We have obtained the aforesaid uncertainty by taking 
the estimate based on MRS and multiplying 
the corresponding event rate by the NNLO K-factor for MRS. 
It may also be mentioned  that the difference between the NLO
estimates of Higgs production using the MRS and CTEQ parameterizations
is rather small ($\lsim 2\%$), according to recent studies
\cite{kfactor}. 
Therefore, it is expected that the  NNLO estimate of uncertainties
(where there is scope of further evolution in any case) used
by us will ultimately converge to even better agreement with other
parameterizations  and will not introduce any serious inaccuracy in our 
conclusions. The programme HDECAY ~\cite{hdecay}, including ${\mathcal
O}(\a_s^2)$ contributions, has been used for Higgs decay computations.

\bt[htb]
\begin{center}
\begin{tabular}{|c|c|c|c|}
\hline\hline
{\bf{Parameter}}& {\bf{Central Value}} &{\bf{Present Uncertainty}}&{\bf{LHC Uncertainty}}\\\hline 
   $m_h$  &   $120.-150.$   &   $-$      &   $0.2 $  \\\hline 
   $m_W$  &   $80.425$      &   $.034$   &   $.015$ \\\hline 
   $m_t$  &   $172.7 $      &   $2.9$    &   $1.5 $  \\\hline  
   $m_b$  &   $4.62  $      &   $.15$    &   $ -  $  \\\hline  
   $m_c$  &   $1.42  $      &   $ .1 $   &   $ -  $  \\\hline  
$m_\tau$  &   $1.777 $      &   $.0003$  &   $ -  $  \\\hline  
$\alpha_s$&   $0.1187$      &   $0.002$  &   $ -  $  \\\hline\hline
\end{tabular}
\caption {\it{Current and projected uncertainties (at LHC) in the values 
of various parameters. All the masses are given in $GeV$. The values
are extracted from refs~\cite{pdg,topmass,param}}}
\label{param.tbl}
\end{center}
\et
The number of two-photon events seen is given by $\cal{L}$$N$ where $\cal{L}$
is the integrated luminosity. $\cal{L}$ is expected to be known at the LHC
to within 2 \%. We include this uncertainty in our calculation, although it 
has a rather small effect on our conclusions. 

The possible sources of theoretical uncertainties can be divided into two
general classes: parametric uncertainties and intrinsic uncertainties.
The former are related to the fact that, within the SM, each quantity of
interest is a function of a set of input parameters, which are known with a
finite experimental precision. Any variation of the input parameters within
the experimentally allowed range gives rise to an uncertainty on the
observable considered.
On the other hand, the intrinsic uncertainties have to do with the
perturbative treatment of the quantum corrections: scheme dependence,
ignorance of higher orders in the perturbative expansion and so on. 
We have included the NNLO K-factors for the production cross-section 
$\sigma ( gg \to h)$, available in the literature and assume that our 
ignorance of more higher order contributions will not introduce a very 
significant uncertainty.

In order to estimate the total uncertainty in $N$, one has to first
obtain the spread in theoretically predicted value in the SM due to the 
uncertainty in the  various parameters used. In addition, however, there 
is an uncertainty in the experimental values, although the actual level 
of this will be known only after the LHC run begins, the anticipated 
statistical spread in the measured value can be estimated through 
simulations. 
These two uncertainties, combined in quadrature, are indicative of the
difference with central value of the SM prediction which
is required to establish any non-standard effect at any given confidence 
level. We have performed such an exercise, taking the standard model
calculation and that with SM + UED contributions.
\bt
\begin{center}
\begin{tabular}{|c|c|c|c|c|c|}
\hline\hline
\multicolumn{6}{|c|}{\bf{Total Uncertainty in SM rate}}\\
\hline
      & & &\multicolumn{3}{ c|}{\bf{PDF+scale uncertainty}}
\\ 
{\bf {Higgs mass~(GeV)}}&$\delta_{exp}(\%)$&$\delta_{th}(\%)$&
                         $(15.0\%)$&$(10.0\%)$&$( 5.0\%)$\\\hline
$120.0$&8.9&8.1&$19.3\%$&$15.8\%$&$13.2\%$\\
\hline  
$130.0$&8.1&6.9&$18.5\%$&$14.7\%$&$11.9\%$\\
\hline  
$140.0$&8.6&5.6&$18.3\%$&$14.4\%$&$11.6\%$\\
\hline  
$150.0$&11.3&4.6&$19.4\%$&$15.9\%$&$13.3\%$\\\hline\hline  
\end{tabular}
\caption{\it{Entries in the second (third) column corresponds to 
experimental (theoretical) uncertainties in the rates as discussed in 
the text. The total uncertainty in the SM rate including the theoretical and
experimental uncertainties along with the errors (different choices) 
due to parton distributions and renormalisation scale (15\%,10\% and 5\%) 
are listed for different Higgs boson mass.}
\label{total.tbl}}
\end{center}
\et
Thus the total uncertainty in $N$ can be expressed as
\bea
{\left(\frac{\d N}{N}\right)}^2 ~=~ {\left(\frac{\d N}{N}\right)}_{\it th}^2 
~+~ {\left(\frac{\d N}{N}\right)}_{\it exp}^2
\eea
where the theoretical component can be further broken up as
\bea
{\left(\frac{\d N}{N}\right)}_{\it th}^2 ~=~ \frac{1}{N^2}\sum_{i}{\s^2_{N_i}}
\eea
where $\s_{N_i}$ stands for the spread in the prediction of 
N due to uncertainty in the $i^{th}$ parameter relevant for the calculation. 
The sum runs over $m_h$, $m_W$, $m_t$, $m_b$, $m_{\tau}$ and  
$m_c$, in addition to the uncertainty in the strong coupling $\alpha_s$.
The spread in the predicted value is predicted in each case by 
random generation of values for each parameter (taken to vary one 
at a time) within the allowed range. Thus we obtain  
$\frac{1}{N^2}{\s^2_{N_i}}$ corresponding to each parameter. This has
been listed in Table~\ref{param.tbl} for different choices of $m_h$.
One has to further include QCD uncertainties arising via 
parameterization dependence of the parton distribution functions (PDF)
\cite{pdfscale} and the renormalisation  scale. Although
NNLO calculation reduced such uncertainties, the net spread in the
prediction due to them could be as large as $\sim 15$ percent
~\cite{pdfscale,kfactor,hocorrs1,hocorrs2,Belyaev,scalevarn} in the Higgs mass 
range $120 -150$ GeV. The levels of uncertainties in the various 
parameters, are presented in 
Table~\ref{param.tbl}. In that table we have given the uncertainties,
wherever they are available, from recent and current experiments like the 
LEP and the Tevatron. In addition, whatever improved measurement, 
leading to smaller errors (in, say, $m_t$ or $m_W$) are expected after 
the initial run of the LHC are also separately incorporated in the table. 
We have used the estimates 
corresponding to LHC wherever they are available.  
In our calculation, we have used three values of the combined uncertainty 
from PDF and scale-dependence, namely, 15\%, 10\% and 5\%, the latter
two with an optimistic view to likely
improvement using data at the LHC. Table 2 contains the finally
predicted values of ${\left(\frac{\d N}{N}\right)}$, for the different
values of the Higgs boson mass. 

\begin{figure}[htb]
\begin{center}
\includegraphics[width=3.05in,height=2.95in]{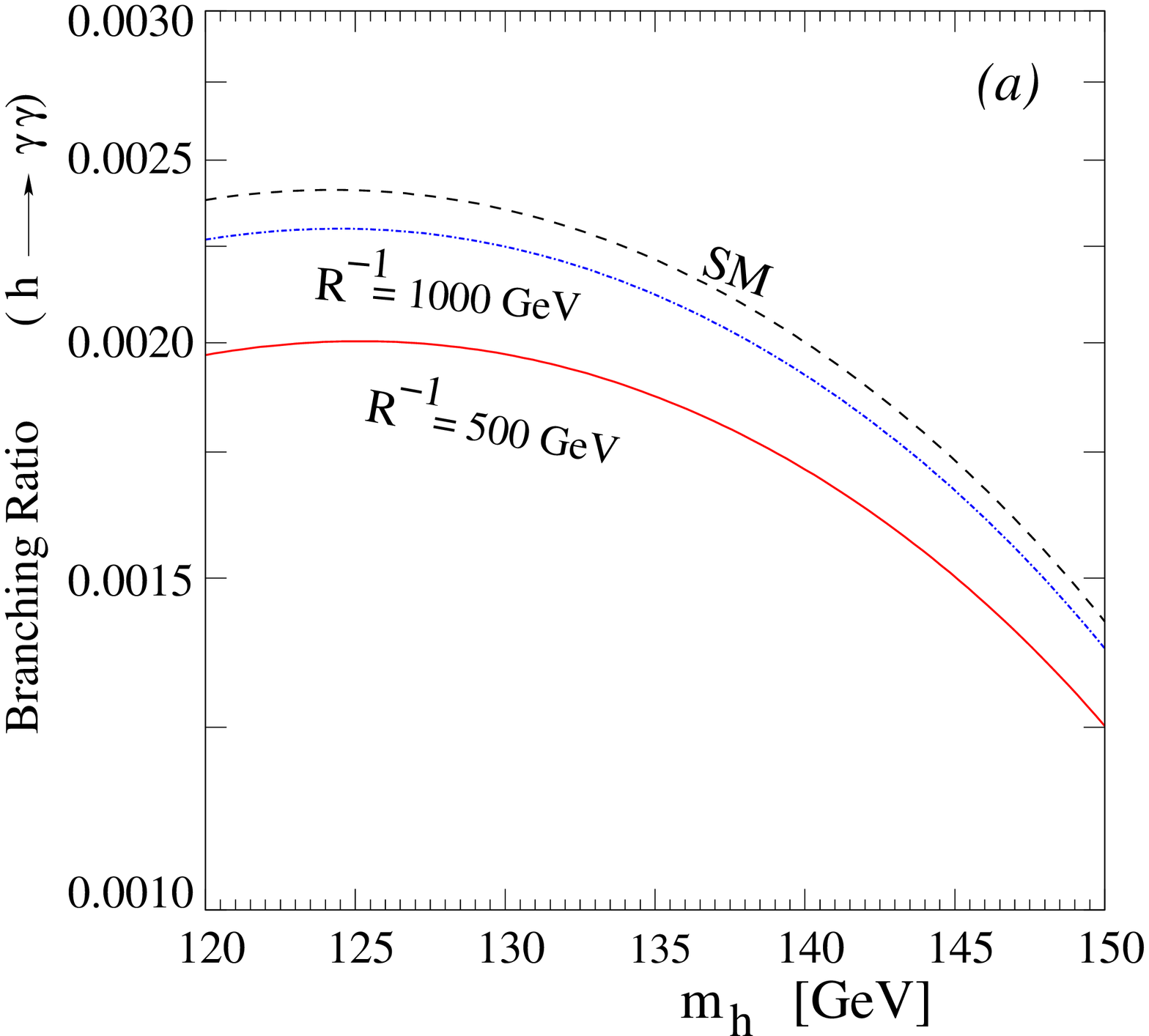}
\includegraphics[width=3in,height=3in]{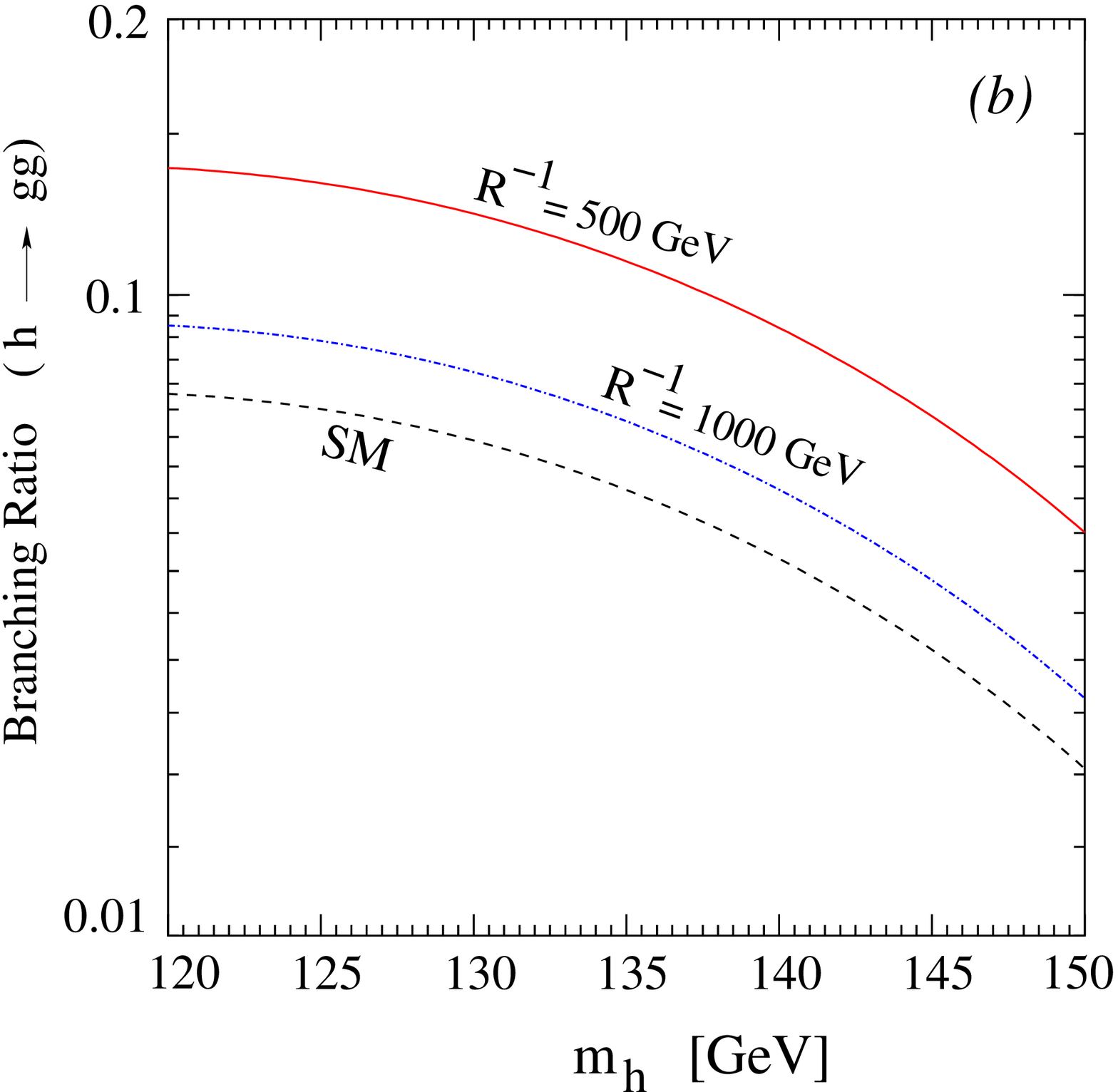}
\caption{ \it Illustrating the effects of UED contributions on the branching
ratio of (a) $h \to \g\g$ and (b) $h \to gg$.}
\label{bratio}
\end{center}
\end{figure}
$N_{\it exp}$ includes  statistical uncertainties, 
as estimated in detector simulations with a luminosity of 100
$fb^{-1}$\cite{expnos}. 
As has been already mentioned, we have  obtained benchmark 
values of this quantity using the results for CMS presented in
ref~\cite{expnos} for 
MRS distributions at the lowest order, and appropriately improving them 
with the NNLO K-factors available in the literature. The resulting 
predictions are  listed as $\delta_{exp}$ in Table~\ref{total.tbl} for
different values of $m_h$. Thus one is able to obtain the 
net ($1 \s$ level) uncertainties in the standard model as shown in the
last three columns of Table~\ref{total.tbl}. 
\begin{figure}[htb]
\begin{center}
\includegraphics[width=2.8in,height=2.8in]{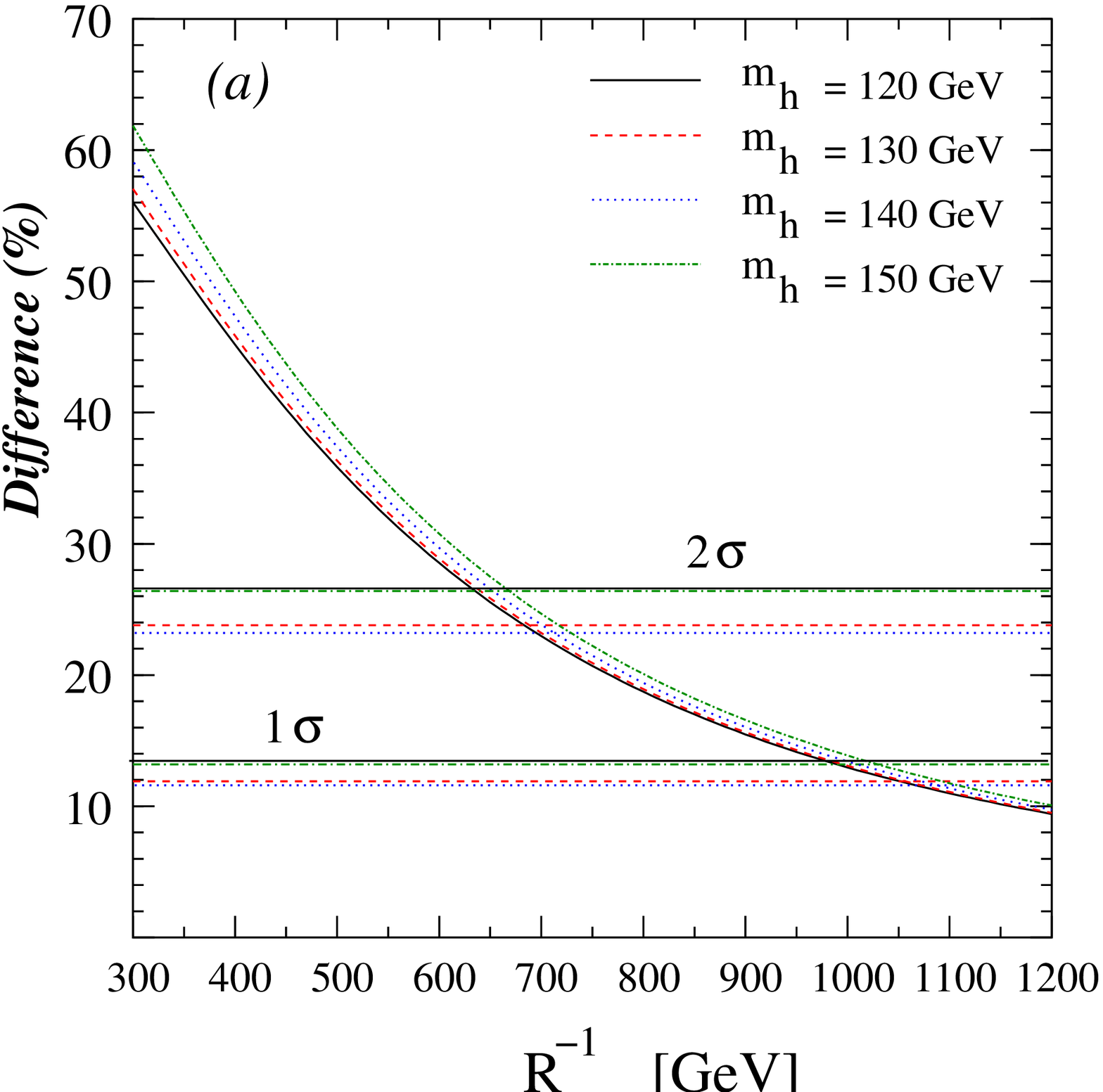}
\includegraphics[width=2.8in,height=2.8in]{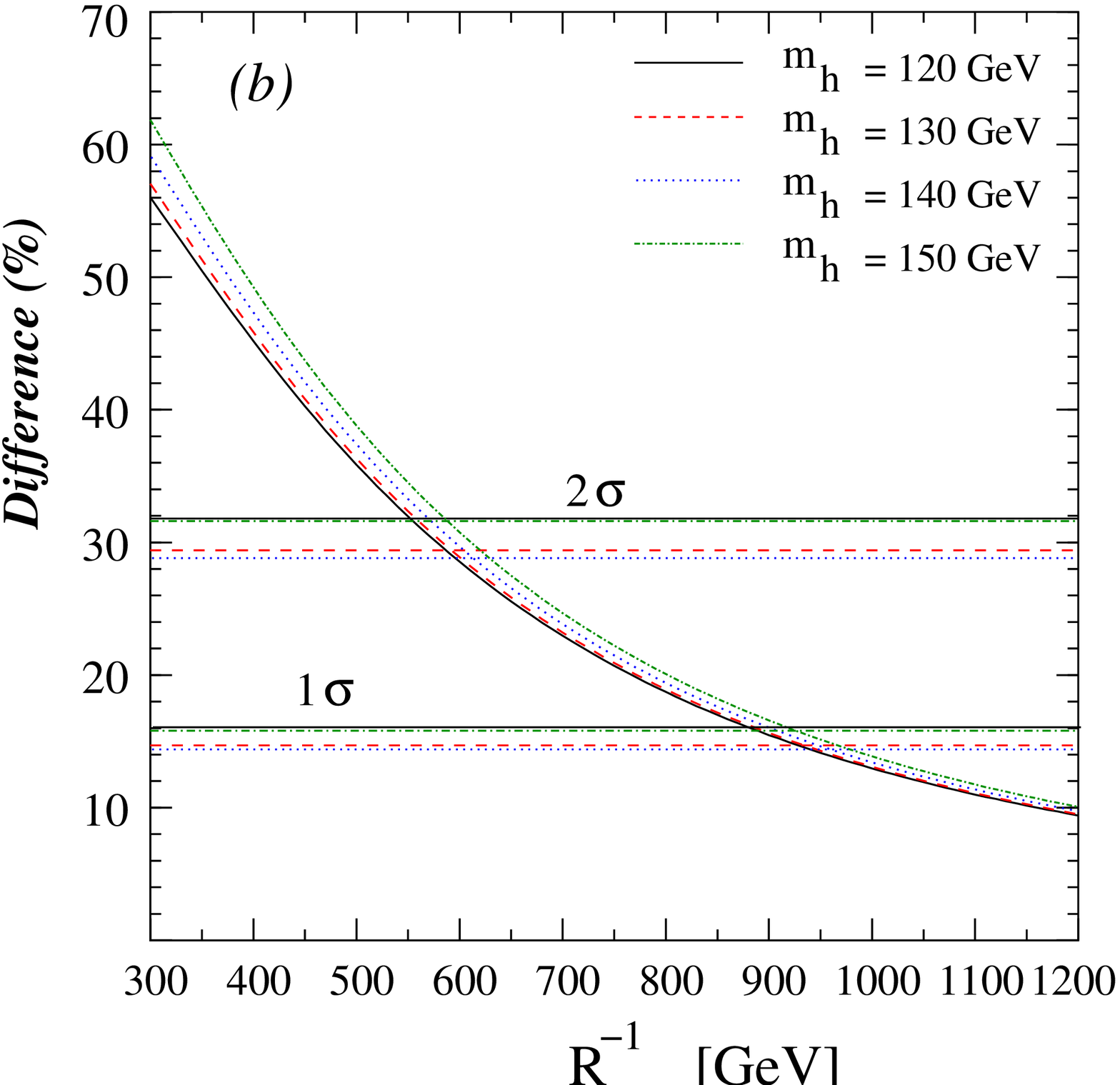}
\includegraphics[width=2.8in,height=2.8in]{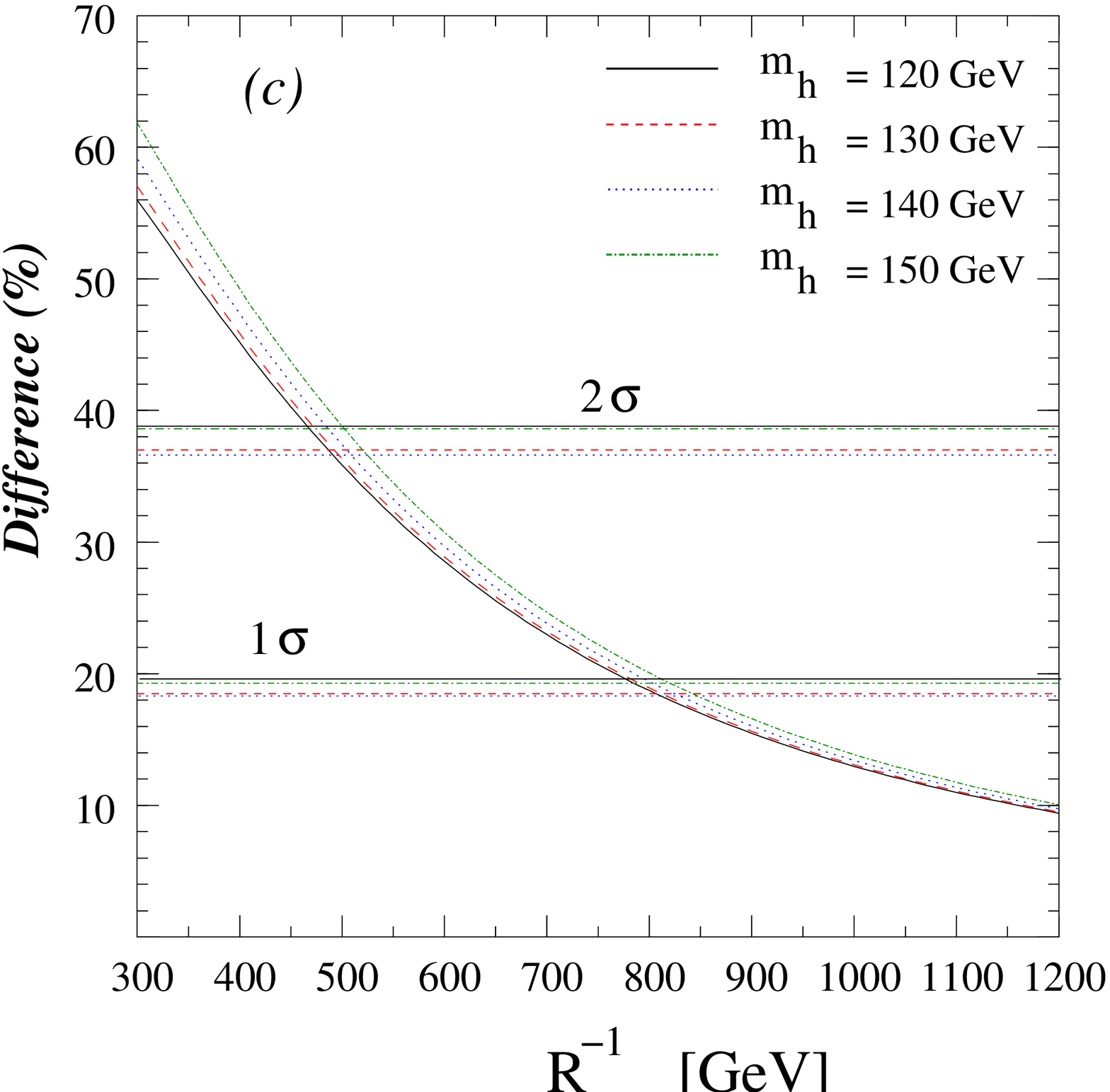}
\caption{\it Illustrating the percentage difference over
standard model rates for different values of the Higgs mass 
as a function of the compactification scale $R^{-1}$. 
The horizontal lines in each figure corresponds to the confidence levels 
as labeled. The graphs are shown for different choices of the 
PDF + scale uncertainty, viz. (a) 5\%, (b) 10\% and (c) 15\%.}
\label{diff}
\end{center}
\end{figure}

Next, the UED contributions via KK excitations-induced diagrams are 
calculated and added to the standard model amplitude. The observable 
decay rate obtained therefrom is compared with that predicted in the 
standard model taking the uncertainty into account at various confidence 
levels. Thus one is able to decide whether the UED contributions to
the diphoton rate are discernible from the standard model contributions 
at a given confidence level for a particular $R$.
The realistic estimate requires subjecting the predictions to
some experimental cuts aimed at maximizing the signal-to-background ratio
as well as focusing on  kinematic regions  of optimal observability.
We incorporate the effects of such cuts with the help of an efficiency
factor which, on explicit calculation in representative cases, 
turns out to be approximately 47-53\%. The only assumptions required are 
that the percentage error due to various parameters are the same for 
uncut rates as those calculated with cuts, and that the standard model 
and UED contribution suffer the same reduction due to cuts. We have 
checked that this holds true so long as the kinematic region is not 
drastically curtailed by the cuts.  

\section{Numerical estimate: discussions}
Our purpose is to see at what confidence levels one can distinguish the
UED effects on $h \longrightarrow \gamma\gamma,~gg$. With this in view,
we estimate the excess in the rates due to the UED contributions and
calculate the fractional difference with that predicted for standard
model. 
It is worth noting that for the mass range of the Higgs boson that 
we consider, the partial decay width of the $h\to\g\g$ mode
falls below the SM value while that of the $h\to gg$ mode is greater
than that of the SM contribution. To highlight the 
dependence, we plot the branching ratios for these two modes in 
figure~\ref{bratio}. This suggests that the UED contribution will have
a slight suppression as the rate for the process in consideration is
proportional to the product of the above branching ratios. 
In figure~\ref{diff} we show the contour plots for the three choices of
PDF + scale uncertainty. We can see from figure~\ref{diff}(a),
which has the most optimistic choice of 5\% for the PDF + scale uncertainty,
that at the 2$\s$ confidence level, one can see excess over the SM rate
for values of compactification scale as large as 
$R^{-1} \simeq 630, 690, 710, 660$ GeV for Higgs mass $m_h = 120,
130, 140, 150$ GeV respectively, while at 1$\s$ confidence level, these
go up to $R^{-1} \simeq 990,1050,1090,1020$ GeV for Higgs mass $m_h
= 120,130, 140, 150$ GeV respectively. For the more conservative
choices of PDF + scale uncertainty, these numbers for $R^{-1}$ will
go down and as shown in figure~\ref{diff}(c), where the choice for PDF
+ scale uncertainty is taken as 15\%, the 2$\s$ (1$\s$) confidence 
level limits $R^{-1} \simeq 450(780),490(810),500(830),500(810)$ GeV 
for Higgs mass $m_h = 120,130,140,150$ GeV respectively. 
With improvements in the measurement resolutions and lower uncertainties, 
this reach can be improved further. We must point out that although the better 
convergence of the NNLO result over the NLO calculations do suggest a better 
understanding of the theoretical result, our ignorance of corrections beyond 
NNLO limits our complete knowledge of the intrinsic theoretical error.  
The updated lower bounds on the compactification scale \cite{Flacke,Gogoladze} 
($\gsim 600$ GeV for $m_h=115$ GeV and top quark mass of 173 GeV at 90\% C.L.),
which depends on the Higgs mass and the top quark mass, however suggest that 
the visible effects at the LHC would be marginal and one really needs a better 
hold on the different uncertainties to highlight the large deviations that 
are expected in the UED predictions. 

\section{Summary and conclusions}
We have explored the signals for the intermediate mass range of the 
Higgs boson production through the $gg \to h$ channel and its
subsequent decay to two photons. Both the production and decay channel
get contributions through loops due to the absence of tree level
couplings. The excited KK modes of the SM particles would give additional
contribution to these decay modes, thus altering the decay rates.
We have performed a detailed analysis of the rates for the inclusive
process $gg \to h \to \g\g$ incorporating the different uncertainties 
that would affect measurements at LHC. We find that one can have 
observable enhancements in the allowed range of the parameter space of UED 
to distinguish effects of UED contributions in the Higgs signals by studying 
the rates for the inclusive process considered in this work. 
We show that although the rates (after including UED contributions) differ 
from the standard model prediction substantially, large uncertainties, both 
theoretical (we have neglected the intrinsic theoretical error in the Higgs 
production cross section from corrections beyond NNLO assuming it to be small) 
as well as experimental, have a large role to play. These 
uncertainties combine to dilute the substantially significant deviations from 
SM coming from new physics contributions and are seen to be marginal 
\cite{Flacke,Gogoladze}. However, with better luminosity and improvements in the 
theoretical calculations and experimental measurements, 
this channel can provide for new physics effects in the Higgs signal itself.  

\bigskip


\end{document}